\documentclass{article}

\usepackage[utf8]{inputenc} 
\usepackage[T1]{fontenc}    
\usepackage{hyperref}       
\usepackage{url}            
\usepackage{booktabs}       
\usepackage{amsfonts}       
\usepackage{nicefrac}       
\usepackage{microtype}      

\usepackage{subfig}
\usepackage{bm}
\usepackage{tikz}
\usetikzlibrary{matrix}
\usepackage{array}
\usepackage{multirow}
\usepackage{verbatim}
\usepackage{amsmath}
\usepackage{amssymb}
\usepackage{amsfonts}
\usepackage{navigator}

\newcommand{\walderurl}{http://bit.ly/1PqNTJ2}

\makeatletter
\newcommand{\thickhline}{%
    \noalign {\ifnum 0=`}\fi \hrule height 1pt
    \futurelet \reserved@a \@xhline
}
\newcolumntype{"}{@{\hskip\tabcolsep\vrule width 1pt\hskip\tabcolsep}}
\makeatother

\newcommand{\chapternewpage}{~ \ifodd\value{page} \chapter*{} \fi}

\newcommand{\novelty}[3]{\ifthenelse{\equal{\value{#1}}{0}}{#3}{#2}\setcounter{#1}{1}}

\newcommand{\etc}{\textit{etc.}}

\newcommand{\beqn}{\begin{equation}\/}
\newcommand{\eeqn}{\end{equation}\/}
\newcommand{\beqns}{\begin{equation*}\/}
\newcommand{\eeqns}{\end{equation*}\/}
\newcommand{\beqna}{\begin{eqnarray}\/}
\newcommand{\eeqna}{\end{eqnarray}\/}
\newcommand{\beqnas}{\begin{eqnarray*}\/}
\newcommand{\eeqnas}{\end{eqnarray*}\/}

\newcommand{\balign}{\begin{align}\/}
\newcommand{\ealign}{\end{align}\/}
\newcommand{\bals}{\begin{align*}\/}
\newcommand{\eals}{\end{align*}\/}

  {\begin{center}\begin{Sbox}\begin{minipage}}%
  {\end{minipage}\end{Sbox}\fbox{\TheSbox}\end{center}}

  {\begin{center}\begin{Sbox}\begin{minipage}}%
  {\end{minipage}\end{Sbox}\Ovalbox{\TheSbox}\end{center}}

\newcommand{\cmcolor}{}

\title{Symbolic Music Data Version 1.0}

\author{
  Christian Walder \\
  CSIRO Data61 \\
  7 London Circuit, Canberra, 2604, Australia. \\
  \texttt{christian.walder@data61.csiro.au} \\
}
\begin{document}
\maketitle

\begin{abstract}
In this document, we introduce a new dataset designed for training machine learning models of symbolic music data. Five datasets are provided, one of which is from a newly collected corpus of 20K midi files. We describe our preprocessing and cleaning pipeline, which includes the exclusion of a number of files based on scores from a previously developed probabilistic machine learning model. We also define training, testing and validation splits for the new dataset, based on a clustering scheme which we also describe. Some simple histograms are included.
\end{abstract}

\section{Introduction}

In this appendix we provide an overview of the symbolic music datasets we offer in pre-processed form\footnote{The data is available for download here: \urllink{\walderurl}{\walderurl} \label{foot:dataurl}}. Note that the source of four out of five of these datasets is the same set of midi files used in \cite{bl}, which also provides pre-processed data. That work provided ``piano roll'' representations, which essentially consist of a regular temporal grid (of period one eighth note) of on/off indicators for each midi note number. While the piano roll is an excellent simplified music format for early investigations into symbolic music modelling, it does have several limitations, as discussed in previous work \cite{beyondthepianoroll}. To name one such limitation, the piano roll format does not explicitly represent note endings, and therefore cannot differentiate between, say, two successive eighth notes, and a single quarter note.

To address these limitations, we have extracted additional information from the same set of midi files. Our goal is to represent the performance (or sounding) of notes by when they begin and end, rather than whether they are sounding or not at each time on a regular temporal grid. The representation we adopt consists of sets of five-tuples of integers representing the:
\begin{itemize}
\item piece number (corresponding to a midi file),
\item track (or part) number, defined by the midi channel in which the note event occurs,
\item midi note number, ranging 0-127 according to the midi standard, and 16-110 inclusive for the data we consider here,
\item note start time, in ``ticks'', (2400 ticks = 1 beat = one quarter note),
\item note end time, also in ticks.
\end{itemize}
This document provides some background on the data, with a special emphasis on our new relatively large dataset, which we derived from an archive kindly provided to us by Pierre Schwob of \urllink{http://www.classicalarchives.com}{http://www.classicalarchives.com}. We are permitted to release this data in the form we provide, but not to provide the original midi files.
%
Please refer to the data archive itself$^\text{\ref{foot:dataurl}}$ for a detailed description of the format. 


A summary of the five datasets is provided in \autoref{table:datasummary}.

\begin{table*}
\begin{center}
\begin{tabular}{ccccc}
\\\thickhline
Dataset & Long Name & Source & Total Pieces & Midi Resolution
\\\thickhline
PMD & \texttt{piano-midi.de} & \cite{poliner,bl} & 124 & 480 \\
\hline
JSB & J.S Bach Chorales & \cite{moray,bl} & 382 & 100 \\
\hline
MUS & MuseData & \cite{musedata,bl} & 783 & 240 \\
\hline
NOT & Nottingham & \cite{nottinghamdata,bl} & 1037 & 480 \\
\hline 
CMA & Classical Midi Archives & \cite{classicalarchives} (new) & 19700 & variable \\
\hline
\end{tabular}
\end{center}
\caption{
\label{table:datasummary}
A summary of the datasets used in this study.
}
\end{table*}
\begin{tabular}{|c|c|c|}
\end{tabular}

\section{Preprocessing}

We applied the following processing steps and filters to the raw midi data.
\begin{itemize}
\item Combination of piano ``sustain pedal'' signals with key press information to obtain equivalent individual note on/off events. 
\item Removal of duplicate/overlapping notes which occur on the same midi channel (while not technically allowed, this still occurs in real midi data due to the permissive nature of the midi file format). Unfortunately, this step is ill posed, and different midi software packages handle this differently. Our approach involves processing notes sequentially in order of start time, and ignoring those note events that overlap a previously added note event.
\item Removal of midi channels with less than two note events (these occurred in the MUS and CMA datasets, and were always information tracks containing authorship information and acknowledgements, \etc).
\item Removal of percussion tracks. These occurred in some of the Haydn symphonies and Bach Cantatas contained in the MUS dataset, as well as in the CMA dataset. It is important to filter these as the percussion instruments are not necessarily pitched, and hence the midi numbers in these tracks are not comparable with those of pitched instruments, which we aim to model.
\item Re-sampling of the timing information to a resolution of 2400 ticks per quarter note, as this is the lowest common multiple of the original midi file resolutions (see \autoref{table:datasummary}) for the four datasets considered in \cite{bl}. We accept some quantization error for some of the CMA files, although 2400 is already an unusually fine grained midi quantization (\textit{cf.} the resolutions of the other datasets, in \autoref{table:datasummary}).
\end{itemize}

\begin{figure*}%
\begin{center}
  \includegraphics[width=0.8\textwidth]{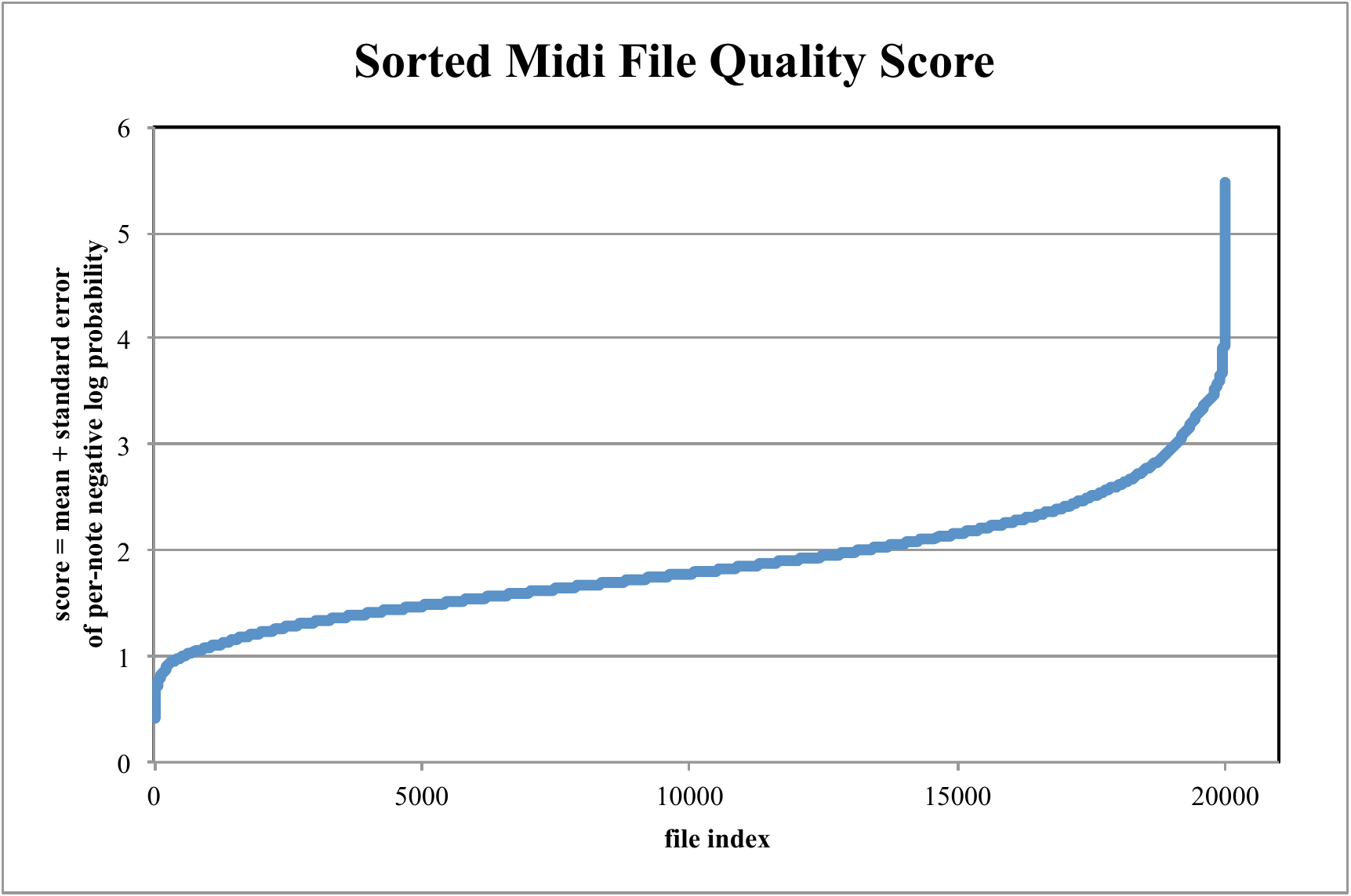}
  \caption{
  \label{fig:scores}
  Our filtering score for the original 20,006 midi files provided by the website \texttt{http://www.classicalarchives.com}. We kept the top 19,700, discarding files with a score greater than 3.9.
  }
\end{center}
\end{figure*}

For our new CMA dataset, we also removed 306 of the 20,006 midi files due to their suspect nature. We did this by assigning a heuristic score to each file and ranking. The score was computed by first training our model \cite{beyondthepianoroll} on the union of the four (transposed) datasets, JSB, PMD, NOT and MUS. We then computed the negative log-probability of each midi note number in the raw CMA data under the aforementioned model. Finally, we defined our heuristic score as, for each file, the mean of these negative log probabilities plus the standard error. The scores we obtained in this way are depicted in \autoref{fig:scores}. A listening test on the best and worst files verified the effectiveness of this measure. In any case, some degree of noise is to be expected in a data set of this size, and should be handled by subsequent modelling efforts.

\section{Splits}

The four datasets used in \cite{bl} retain the original training, testing, and validation splits used in that work. For CMA, we took a careful approach to data splitting. The main issue was data duplicates, since the midi archive we were provided contained multiple versions of several pieces, each encoded slightly differently by a different transcriber. To reduce the statistical dependence between the train/test/validation splits of the CMA set, we used the following methodology:
\begin{enumerate}
\item We computed a simple signature vector for each file, which consisted of the concatenation of two vectors. The first was the normalised histogram of midi note numbers in the file. For the second vector, we quantized the event durations into a set of sensible bins, and computed a normalised histogram of the resulting quantised durations.
\item Given the signature vectors associated with each file, we performed hierarchical clustering using the function \texttt{scipy.cluster.hierarchy.dendrogram} from the python scipy library\footnote{\urllink{https://www.scipy.org}{https://www.scipy.org}}. We then ordered the files by traversing the resulting hierarchy in a depth first fashion.
\item Given the above ordering, we took contiguous chunks of 15,760, 1,970 and 1,970 files for the train, test, validation sets, respectively. This leads to a similar ratio of split sizes as in \cite{bl}.
\end{enumerate}

\section{Basic Exploratory Plots}

We provide some basic exploratory plots in figures \ref{fig:PMD}--\ref{fig:MUS}. 

The \textbf{Note Distribution} and \textbf{Number of Notes Per Piece} plots are self explanatory.

Note that the \textbf{Number of Parts Per Piece} (lower left sub figure) is fixed at one for the entire JSB dataset. This is due to an unfortunate lack of midi track information in those files, many of which are in fact four part harmonies. The pieces in the NOT dataset feature either one part (in the case of pure melodies) or two (in the case of melodies with associated chord accompaniments). The PMD dataset features up to six parts (for a three-part Bach fugue in which left and right hands are tracked separately). MUS features up to 27 parts (for Bach's \textit{St. Matthew's Passion}). The CMA data features two pieces with 46 parts --- Ravel's \textit{Valses Nobles et Sentimentales}, and \textit{Venus}, by Gustav Holst.

The least obvious sub-figures are those on the lower-right labeled \textbf{Peak Polyphonicity Per Piece}. Polyphonicity simply refers to the number of simultaneously sounding notes, and this number can be rather high. For the PMD data, this is mainly attributable to musical ``runs'' which are performed with the piano sustain pedal depressed, for example in some of the Liszt pieces. For the MUS data, this is mainly due to the inclusion of large orchestral works which feature many instruments. The CMA data, of course, contains both of the aforementioned sources of high levels of polyphonicity.

\section*{Acknowledgements}

Special thanks to Pierre Schwob of \urllink{http://www.classicalarchives.com}{http://www.classicalarchives.com}, who permitted us to release the data in the form we describe.

\bibliography{walder}
\bibliographystyle{hep}

\newcommand{\dataplot}[3]{%
\begin{figure*}%
\begin{center}%
  \includegraphics[width=0.45\textwidth,page=1]{#1}%
  \includegraphics[width=0.45\textwidth,page=2]{#1}%
  \\
  \includegraphics[width=0.45\textwidth,page=3]{#1}%
  \includegraphics[width=0.45\textwidth,page=4]{#1}%
  \caption{%
  \label{fig:#2}%
  Summary of the #2 dataset. #3%
  }%
\end{center}%
\end{figure*}%
}

\dataplot{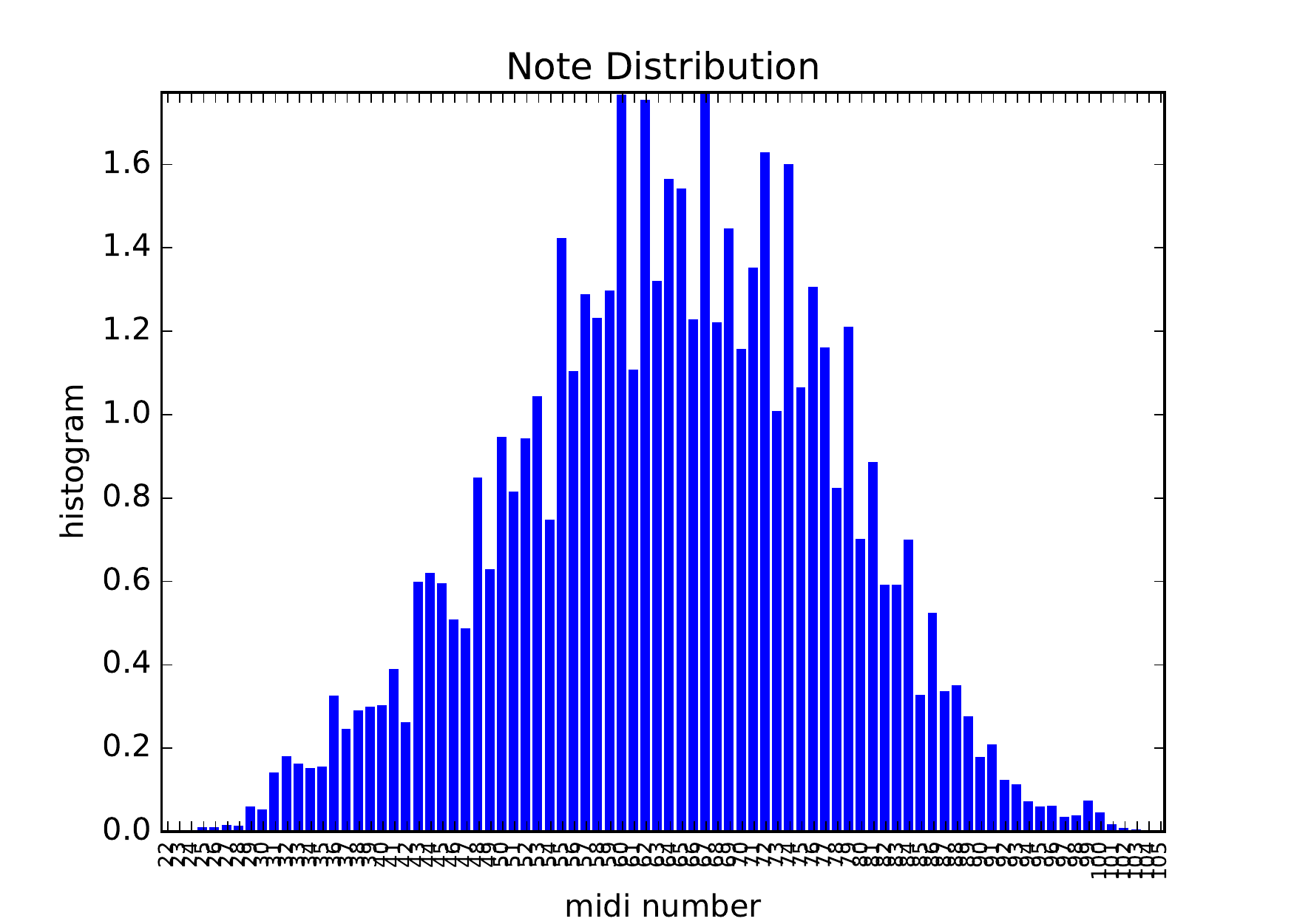}{PMD}{}
\dataplot{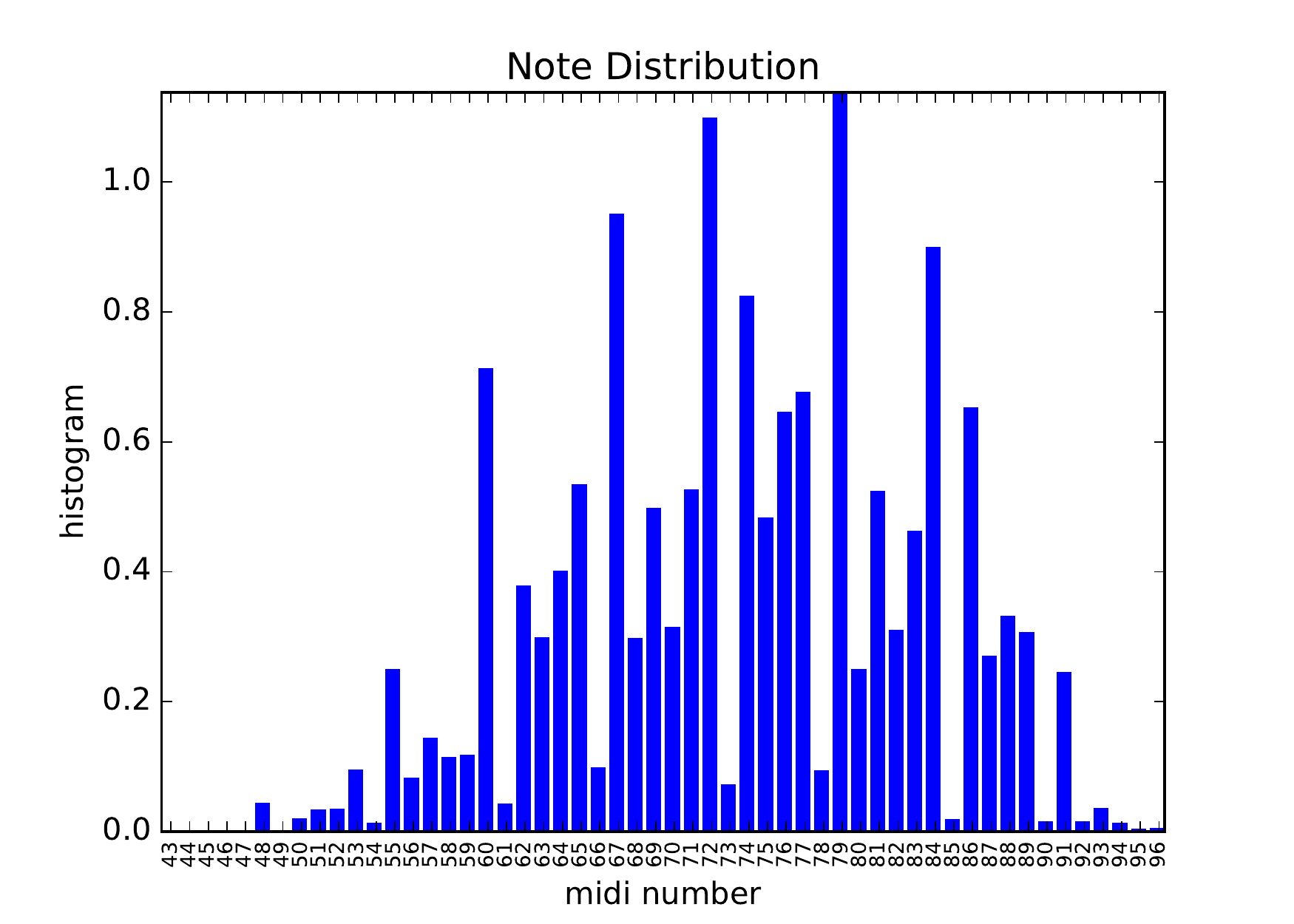}{JSB}{}
\dataplot{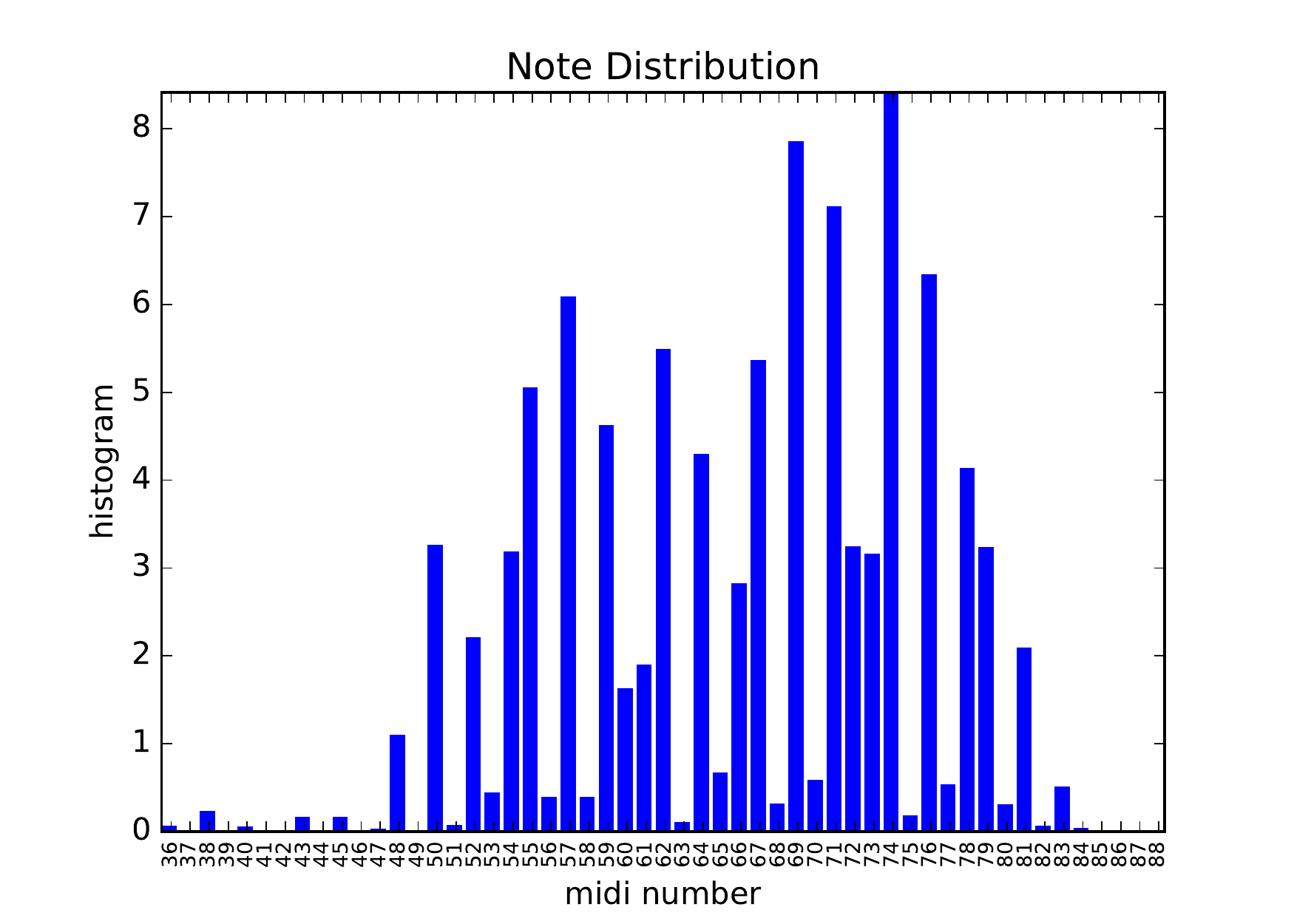}{NOT}{}
\dataplot{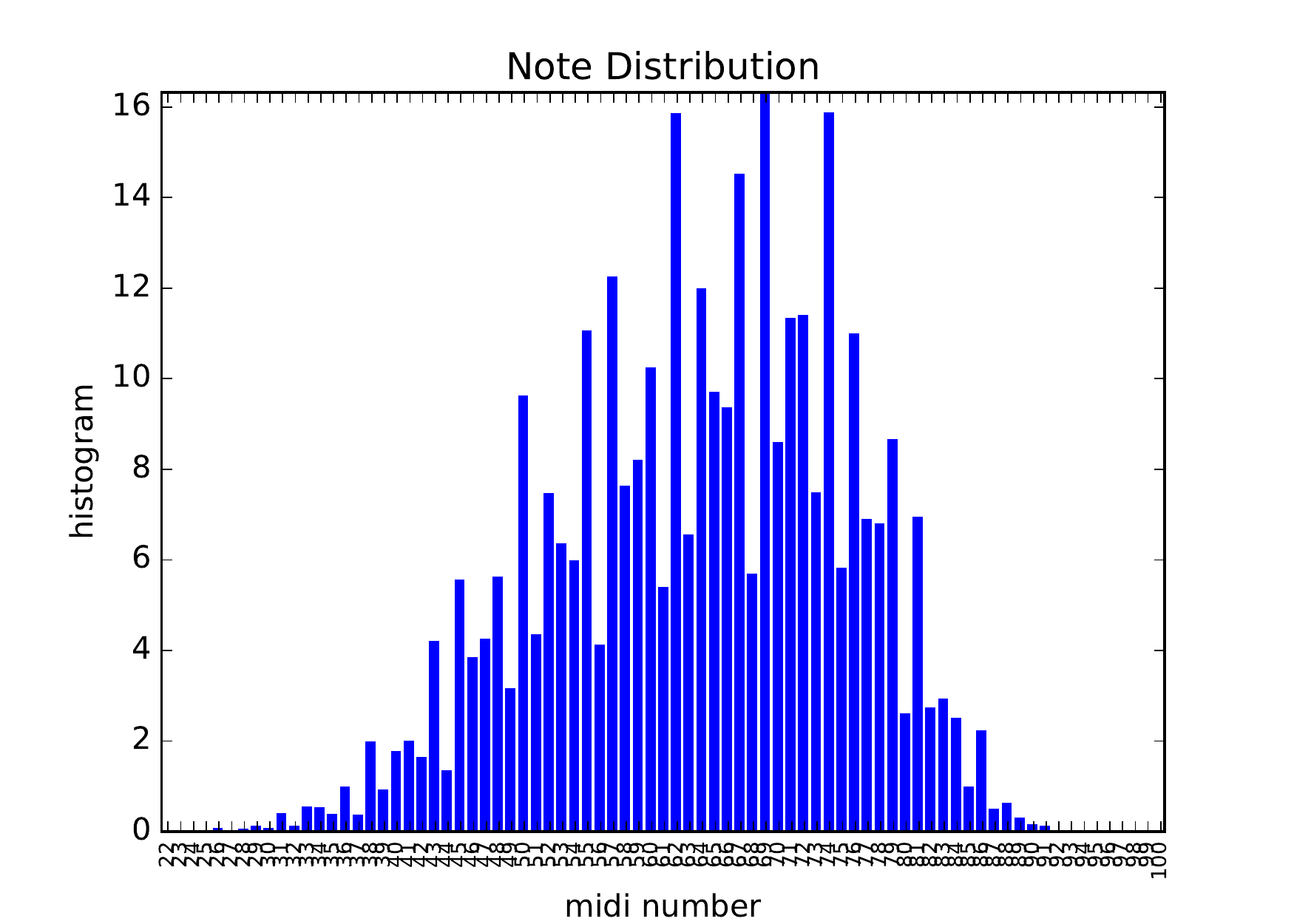}{MUS}{}
\dataplot{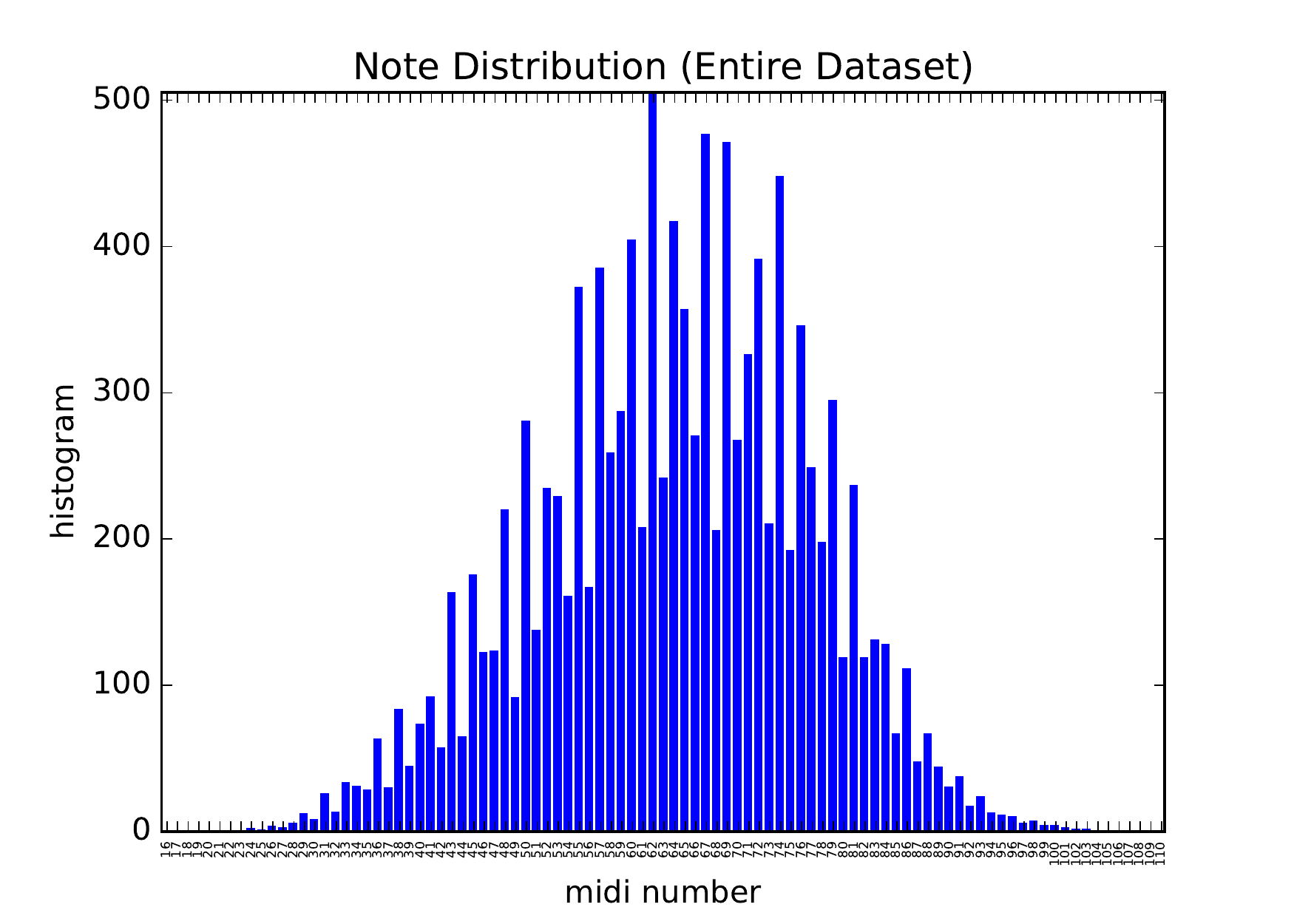}{CMA}{Note the log scale on three of the plots.}
 
\end{document}